\documentclass[12pt]{iopart}
\usepackage{dcolumn}
\usepackage{amssymb}
\usepackage{graphicx}
\usepackage{float}
\usepackage{multirow}
\usepackage{subfigure}
\usepackage{bm}

\begin{document}
\title{Can a Lamb Reach a Haven Before Being Eaten by Diffusing Lions?}

\author{Alan Gabel$^1$, Satya N. Majumdar$^2$, Nagendra K. Panduranga$^1$, and S. Redner$^1$}
\address{$^1$Center for Polymer Studies and Department of
  Physics, Boston University, Boston, MA 02215, USA}
\address{$^2$Laboratoire de Physique Th\'eorique et
  Mod\`eles Statistiques (UMR 8626 du CNRS), Universit\'é Paris-Sud,
  B\^atiment 100, 91405 Orsay Cedex, France}

\begin{abstract} 
  We study the survival of a single diffusing lamb on the positive half line
  in the presence of $N$ diffusing lions that all start at the same position
  $L$ to the right of the lamb and a haven at $x=0$.  If the lamb reaches
  this haven before meeting any lion, the lamb survives.  We investigate the
  survival probability of the lamb, $S_N(x,L)$, as a function of $N$ and the
  respective initial positions of the lamb and the lions, $x$ and $L$.  We
  determine $S_N(x,L)$ analytically for the special cases of $N=1$ and
  $N\rightarrow\infty$.  For large but finite $N$, we determine the unusual
  asymptotic form whose leading behavior is $S_N(z)\sim N^{-z^2}$, with
  $z=x/L$.  Simulations of the capture process very slowly converge to this
  asymptotic prediction as $N$ reaches $10^{500}$.
\end{abstract}
\pacs{02.50.Cw, 05.40.-a, 05.50.+q}

\maketitle

\section{Introduction}

We investigate the one-dimensional diffusive capture process in which a
marked particle --- a ``lamb'' --- diffuses on the positive half line $x>0$
in the presence of $N$ independently diffusing predators --- ``lions'' ---
that are all initially at $L>x$.  If the lamb meets any lion, the lamb is
killed.  Additionally, the origin is a \emph{haven} for the lamb.  If the
lamb reaches the haven before meeting any of the lions, then the lamb
survives.  We are interested in the survival probability of the lamb as a
function of the starting positions of the two species, as well as on the
number of lions.

This model is a natural counterpoint to the well-studied capture process of a
single diffusing lamb in the presence of $N$ independent, diffusing lions on
the infinite line~\cite{Br91,Ke92,Kr96,Re99,KMR10}.  In the most interesting
situation where the lions are all on one side of the lamb, the survival
probability $\mathcal{S}_N(t)$ of the lamb asymptotically decays as a
power-law in time, $\mathcal{S}_N(t)\sim t^{-\beta_N}$, with the exponent
$\beta_N$ exhibiting a non-trivial dependence on the number of lions $N$ and
also on the diffusivities of each animal.  For simplicity, the case where the
diffusivities of all animals are the same (and set to one) is normally
considered.  The initial positions of the lamb and the lions are irrelevant
in this asymptotic behavior.

For this diffusive capture on the infinite line, the exponent $\beta_N$ is
known exactly only for $N=1$ and $N=2$: $\beta_1=\frac{1}{2}$ and
$\beta_2=\frac{3}{4}$~\cite{Br91,Ke92,Kr96,Re99,KMR10,Fi84,FG88}.  The latter
result shows that even though the two lions are independent, their effect on
the capture process is not, since $t^{-\beta_2}>\left(t^{-\beta_1}\right)^2$.
For the case $N=3$, a mapping to an equivalent electrostatic problem leads to
the accurate estimate $\beta_3=0.91342\pm 0.00008$~\cite{AJMKR}.  For $N>3$,
the value of $\beta_N$ has been estimated with moderate accuracy only for a
few values of $N$~\cite{Br91}; however, it is known that $\beta_N>1$ for
$N>3$, so that the average lifetime of the lamb is finite~\cite{LS02}.
Because $\beta_N$ grows more slowly than linearly with $N$, each additional
lion has a progressively weaker influence on the capture process.  As
$N\to\infty$, both asymptotic and rigorous arguments give
$\beta_N\to\frac{1}{4}\ln N$~\cite{Ke92,Kr96,Re99}.  Parenthetically, the
capture process with lions sited on both sides of the lamb is much more
efficient than in the one-sided system.  For $N$ lions with approximately
equal numbers of them on either side of the lamb, the lamb survival
probability asymptotically decays as $t^{-\gamma_N}$, with $\gamma_N$ growing
linearly with $N$ for large $N$.

\begin{figure}[htb]
\center{
\includegraphics[width=0.6\textwidth]{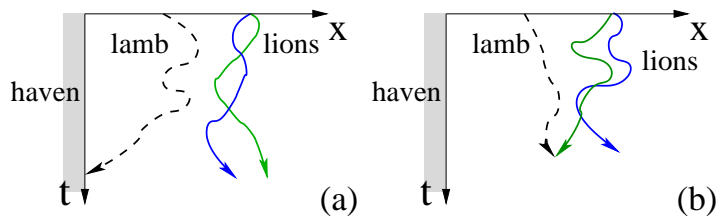}
\caption{Space-time trajectories of a lamb and two lions with a haven at
  $x=0$ when: (a) the lamb survives and (b) the lamb is eaten.}
\label{lL}
}
\end{figure}

In this work, we incorporate the new feature of a \emph{haven} at $x=0$ and
ask whether the lamb can reach the haven before meeting any of the lions.  If
the haven is reached, we say that the lamb survives (Fig.~\ref{lL}).  Our
goal is to determine how the ultimate survival probability $S_N(x,L)$ depends
on the initial positions of the lamb and all the lions, $x$ and $L>x$,
respectively, as well as on the number of lions.  As we shall see, the
survival probability depends on $z\equiv\frac{x}{L}$ rather than on $x$ and
$L$ separately and thus we write the ultimate survival probability as
$S_N(z)$.  Our main result is that $S_N(z)$ has an unusual form whose leading
behavior is $S_N(z)\sim N^{-z^2}$, but this behavior does not become apparent
until $N$ becomes of the order of $10^{500}$.

We begin by solving the simplest and exactly-soluble case of one lion in
Sect.~\ref{exact}.  We also outline the formal solution to the problem for
any number of lions. In Sect.~\ref{inf} we treat the extreme case where the
number of lions is infinite, so that the lion that is closest to the lamb
moves ballistically.  We then investigate arbitrary $N$ in
Sect.~\ref{general}.  When $N$ is large, we can replace the $N$ lions by a
single ``closest lion'' that moves deterministically.  We develop
approximation schemes to estimate $S_N(z)$ in this large-$N$ limit.  We also
present numerical results for the survival probability in Sect.~\ref{sim}.  A
straightforward simulation of the random-walk motion of the particles is
prohibitively slow when $N$ is large, and we present two alternative
approaches that are considerably more efficient and allow us to probe the
survival probability in the regime where $N$ is extremely large --- of the
order of $10^{500}$.  Finally, in Sect.~\ref{conc}, we summarize and also
discuss some natural and intriguing extensions of the model.

\section{Exact Analysis}
\label{exact}

\subsection{One Lion}
\label{one}

As a preliminary, we can readily solve the case of one lamb at $x= x_1$ and
one lion at $L= x_2>x_1$ for the general situation where the diffusivities of
the two species are distinct --- $D_1$ for the lamb and $D_2$ for the lion.
We compute the survival probability that the lamb reaches the haven at $x=0$
before being eaten by the lion, $S(x_1,x_2)$, by mapping the coordinates of
the lamb and the lion on the line to diffusion in a two-dimensional wedge,
from which the survival probability follows easily.

\begin{figure}[ht]
\center{
\includegraphics[width=0.4\textwidth]{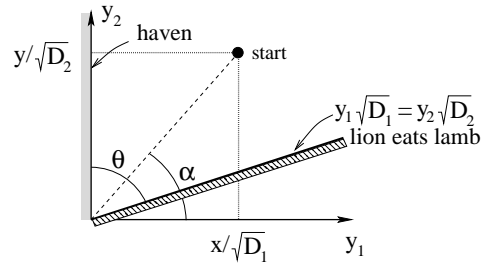}
\caption{Mapping the diffusion of a lamb and a lion on the half line $x>0$ to
  isotropic diffusion in a wedge of opening angle $\theta$. }
\label{wedge}
}
\end{figure}

It is convenient to transform from the coordinates $(x_1,x_2)$ to
$y_1=x_1/\sqrt{D_1}$ and $y_2=x_2/\sqrt{D_2}$.  In the $y_1$-$y_2$ plane, the
motions of the lamb and lion on the half line can be viewed as the isotropic
diffusion of a fictitious composite particle with unit
diffusivity~\cite{FG88,fpp}.  If $y_1$ reaches zero while the condition
$y_1<y_2$ is always satisfied, the lamb survives (Fig.~\ref{wedge}).
Conversely, if $y_1\sqrt{D_1}=y_2\sqrt{D_2}$ at some time (corresponding to
$x_1=x_2$) while $y_1$ always remains positive, then the lamb has been eaten
by the lion before the haven is reached.

In the $y_1$-$y_2$ plane, the initial position of the composite particle is
\begin{equation*}
\left(\frac{x_1}{\sqrt{D_1}},\frac{x_2}{\sqrt{D_2}}\right)~, \nonumber
\end{equation*}
corresponding to the polar angle
\begin{equation}
\alpha =\tan^{-1}\left(\frac{x_2/\sqrt{D_2}}{x_1/\sqrt{D_1}}\right)~.
\end{equation}
The allowed region for the composite particle is a wedge of opening angle
\begin{equation}
\label{theta}
\theta=\tan^{-1}\sqrt{D_2/D_1}\,.
\end{equation}
We want the probability $S(x_1,x_2)$ that the composite particle first hits
the line $y_1=0$ (corresponding to the lamb reaching the haven) without
hitting the line $y_1\sqrt{D_1}=y_2\sqrt{D_2}$.  This probability satisfies
the Laplace equation~\cite{fpp}
\begin{equation*}
  D_1\, \frac{\partial^2 S}{\partial {x_1}^2}+ D_2\,\frac{\partial^2
    S}{\partial {x_2}^2}=0
\end{equation*}
for $x_2\ge x_1$, with boundary conditions $S(x_1\!=\!0, x_2)=1$ and $S(x_1,
x_2\!=\!x_1)=0$.  Clearly the solution is a function that linearly
interpolates between 0 and 1 in the angular direction, so that the ultimate
survival probability is~\cite{fpp,CJ59}
\begin{equation}
  S(x_1,x_2)=\frac{\alpha -(\pi/2-\theta)}{\theta}~.
\end{equation}

\begin{figure}[ht]
\center{
\includegraphics[width=0.4\textwidth]{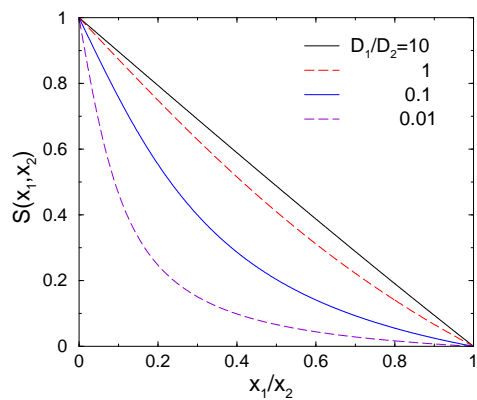}
\caption{Ultimate survival probability $S(x_1,x_2)$ as a function of
  $x_1/x_2$ for various values of the lamb and lion diffusivities, $D_1$
  and $D_2$, respectively.}
\label{S2}
}
\end{figure}

As is obvious from Fig.~\ref{S2}, the closer that the lamb starts to the
haven the more likely it is to survive.  Moreover, as can be inferred from
Fig.~\ref{wedge}, the best strategy for the lamb to survive for a given
initial condition is to diffuse quickly.  As the diffusivity of the lamb
$D_1$ increases, the wedge angle in Fig.~\ref{wedge} approaches
$\frac{\pi}{2}$ while the starting position of the fictitious particle in the
plane moves close to the $y_2$ axis, i.e., closer to the haven.  Finally
notice that in the limit $D_2\to 0$ (stationary lion), the survival
probability decays linearly with $x_1/x_2$.

As a byproduct of the wedge mapping, we can immediately determine the
probability that the lamb is still diffusing --- that is, the lamb has not
yet reached the haven and has not yet been eaten by the lion.  This situation
corresponds to the fictitious particle having not yet reached either of the
sides of an infinite wedge defined by $x_1=0$ and $x_1=x_2$.  In the
isotropic $y_1$-$y_2$ coordinates, this wedge has opening angle $\theta$
(Fig.~\ref{wedge}), and the survival probability asymptotically decays as
$t^{-\pi/2\theta}$.  In particular, when $D_1=D_2$, then
$\theta=\frac{\pi}{4}$ (see Eq.~(\ref{theta})), and the survival probability
asymptotically decays as $t^{-2}$.

\subsection{Formal Solution for General $N$}

The reasoning given above can be readily generalized to map the problem of a
diffusing lamb in the presence of $N$ diffusing lions to a single diffusing
fictitious particle in $N+1$ dimensions, with boundary conditions that
reflect the lamb reaching the haven or being eaten by a lion.  For
simplicity, we set the diffusivities of the lamb and the lions to one.  We
first discuss the case of two lions; the generalization to any number of
lions is immediate.

Suppose that the lamb is initially at $x_1>0$ and that the two lions are
initially at $x_2=x_3>x_1$.  The lamb survives if it reaches $x=0$ without
meeting either of the lions on the way to $x=0$.  We now map the diffusion of
the three interacting particles on the positive half line to the isotropic
diffusion of a composite particle at $(x_1,x_2,x_3)$ in three dimensions,
with constraints that correspond to the interactions in the lamb-lion system.
By this mapping, the allowed region for the composite particle is defined by
$x_1>0$, corresponding to the lamb not yet reaching the refuge, as well as by
$x_1<x_2$ and $x_1<x_3$, corresponding to the lamb not yet eaten by either of
the lions.  This defines a wedge-shaped region that are delineated by three
planar sides that is known as a Weyl chamber~\cite{G99}.

The survival of the lamb corresponds to the composite particle first hitting
the plane $x_1=0$ of the Weyl chamber without hitting either of the planes
$x_1=x_2$ and $x_1=x_3$.  By the equivalence between first-passage and
electrostatics~\cite{fpp}, this survival probability of the lamb coincides
with the electrostatic potential $\Phi(x_1,x_2,x_3)$ at the initial point of
the composite particle, with the boundary conditions $\Phi=1$ on the plane
$x_1=0$, and $\Phi=0$ on the planes $x_1=x_2$ and $x_1=x_3$.  This same
mapping works for any number of lions and constitutes the formal solution.
Unfortunately, the analytical solution to this potential problem does not
seem tractable for more than one lion (i.e., three or more particles),
although some extreme value electrostatic properties have recently been
exactly solved for the three-particle problem~\cite{MB10}.

\section{Infinite Number of Lions}
\label{inf}

When the number of lions is infinite, the lion closest to the haven --- the
closest lion --- would reach the haven at an infinitesimal time.  However, it
is instructive to consider the related problem in which each lion undergoes a
nearest-neighbor random walk.  In this case, the position of the last lion
inexorably moves one lattice spacing to the left in each time step.  For this
system, we determine the ultimate survival probability $S_\infty(x,L)$ by
writing the backward Kolmogorov equation~\cite{fpp} for $S_\infty(x,L)$ and
then applying scaling to solve this equation.  The result should correspond
to that obtained for diffusing lions in the limit of very large $N$.

To write the backward equation, we consider the evolution of the system over
a small time interval $[0,\Delta t]$ during which the lamb moves to
$x+\eta(0)\Delta t$ and the boundary moves to $L - v \Delta t$, where $v$ is
the boundary velocity.  That is, the position of the lamb $x(t)$ evolves by
the Langevin equation $dx/dt=\eta(t)$, where $\eta(t)$ is Gaussian white
noise with zero mean, $\langle \eta(t)\rangle=0$, and correlation $\langle
\eta(t)\eta(t')\rangle= 2\,D\,\delta(t-t')$.  We now view the new positions
of the lamb and the boundary after the time interval $\Delta t$ as the
initial conditions for the subsequent evolution.  Thus $S(x,L)= \langle
S(x+\eta(0) \Delta t, L-v\Delta t)\rangle$, where the average is over the
initial noise $\eta(0)$.  Expanding the right-hand side of this recursion to
lowest non-vanishing order in each variable and using the properties of
delta-correlated noise, we obtain the backward equation
\begin{equation}
  D \frac{\partial ^2S}{\partial x^2} -  v\frac{\partial
    S}{\partial L} =0 
\end{equation}
for $0\le x\le L$, with the boundary conditions $S(0,L)=1$ and $S(L,L)=0$.
To solve this equation we make the scaling ansatz $S(x,L) = f(y)$ (with
$y={x}/{\sqrt{L}}$) to give the ordinary differential equation for $f$:
\begin{equation}
\label{finf}
f'' + \frac{v}{2D}\, y\, f' =0\,,
\end{equation}
subject to the boundary conditions $f(0)=1$ and $f(\sqrt{L})=0$; here the
prime denotes differentiation with respect to $y$.

Integrating and applying the boundary conditions gives
\begin{equation}
\label{fz}
f(y)=1 -
\frac{\mathrm{erf}\big(y\sqrt{{v}/{4D}}\big)}{\mathrm{erf}\big(\sqrt{{vL}/{4D}}\big)}~.
\end{equation}
In the limit $L\to\infty$, this expression reduces to 
\begin{equation}
\label{fza}
f(y)\to\mathrm{erfc}\big(y\sqrt{{v}/{4D}}\big)=\mathrm{erfc}\big(z\sqrt{{vL}/{4D}}\big)\,,
\end{equation}
with $z=x/L$.  The primary feature of this result is that the lamb survival
probability is non-zero only within a thin boundary layer where the starting
position satisfies $x\ll\sqrt{4DL/v}$.  Outside this layer the lamb is almost
surely eaten by one of the lions.

\section{Asymptotics for Large $N$}
\label{general}

\begin{figure}[ht]
\center{
\includegraphics[width=0.3\textwidth]{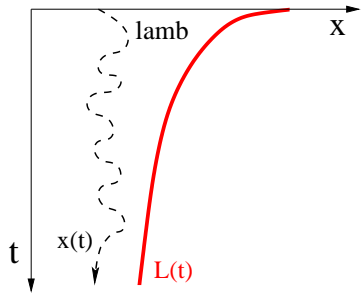}
\caption{Schematic space-time representation of the diffusion of the lamb at
  $x$ and the position $L(t)$ of the closest lion when the number of lions
  $N\gg 1$.}
\label{xlast}
}
\end{figure}

The capture process also simplifies when the number of lions $N$ is finite
but large, because the position of the closest lion becomes progressively
more deterministic as $N$ increases, even though each individual lion
undergoes independent Brownian motion.  Thus we only need to consider the
ultimate survival of the lamb in the presence of a single effective
predator~\cite{movingm} --- the closest lion --- that moves systematically
towards the lamb (Fig.~\ref{xlast}).  We now exploit this physical picture to
give a heuristic argument for the ultimate survival probability of the lamb.

When all the lions start at $L>x$, the average number of lions at $x$ is
\begin{equation*}
n(x,t)=\frac{N}{\sqrt{4\pi Dt}}\,\, e^{-(x-L)^2/4Dt}~.
\end{equation*}
We estimate the location of the closest lion, $L(t)$, by demanding that
$n(L(t),t)=1$.  This criterion gives~\cite{Re99}
\begin{equation}
\label{lstar}
  L(t)= L-\sqrt{A t}\,, 
\end{equation}
where
\begin{equation}
A=4D{\ln M}\left(1-\frac{1}{2}\frac{\ln\ln M}{\ln M}+\ldots\right)
\end{equation}
and $M=N/\sqrt{4\pi}$.  Thus to lowest order, $A\approx4D\ln N$.  At a
critical time $t^*={L^2}/A$ the closest lion has reached the haven at $x=0$
and the capture process is necessarily finished --- either the lamb has been
killed or it has reached the haven.  Notice that although $N$ must be large
for the closest lion to move deterministically, $N$ cannot be too large.  As
discussed in the previous section, if each lion undergoes a nearest-neighbor
random walk, the closest lion moves deterministically to the left with speed
$v=1$ when $N$ is sufficiently large.  For Eq.~(\ref{lstar}) to be valid, we
therefore require (to lowest order) that $\sqrt{4Dt\,\ln N}<vt$, or
$N<\exp(v^2t/4D)$.  Using $v=1$ and $D=\frac{1}{2}$ for a nearest-neighbor
random walk, the last lion moves deterministically as $\sqrt{4Dt\,\ln N}$
only when $t>2\ln N$.  For $t<2\ln N$, the last lion moves with constant unit
speed toward the lamb.

We now crudely estimate the ultimate survival probability of the lamb as the
total probability flux that reaches $x=0$ up to time $t^*$ in the
semi-infinite system \emph{without\/} any additional constraints.  This
integrated flux represents an upper bound for the survival probability for
large $N$ because this estimate includes lamb trajectories that could
intersect the trajectory of the last lion and then reach the haven.  For a
diffusing particle that starts at $x$, the flux to an absorbing boundary at
the origin at time $t$ is:~\cite{fpp}
\begin{equation*}
j(0,t)=\frac{x}{\sqrt{4\pi Dt^3}}\,\, e^{-x^2/4Dt}~.
\end{equation*}
Consequently, the probability $S_N$ for the lamb to get trapped at the origin
up to time $t^*$ (corresponding to the lamb reaching the haven and surviving)
satisfies the bound
\begin{equation}
\label{PN-int}
S_N<\int_0^{t^*} \frac{x}{\sqrt{4\pi Dt^3}}\,\, e^{-x^2/4Dt}\,dt\,
=\mathrm{erfc}(z\sqrt{\ln N}).
\end{equation}
Here we have used the substitution $u=x/\sqrt{4Dt}$ to transform to a
Gaussian integral, as well as the lowest-order approximation
$t^*={L^2}/({4D\ln N})$ and $z=\frac{x}{L}$.

From the asymptotic form $\mathrm{erfc}(x)\sim e^{-x^2}/(\sqrt{\pi\, x^2})$,
we thus obtain an upper bound for the ultimate survival probability that has
the unusual functional form
\begin{equation}
\label{PN-asymp}
S_N< \frac{1}{\sqrt{\pi}}\,\,N^{-z^2}\,\big[{\ln(N^{z^2})}\big]^{-1/2}\,.
\end{equation}
Consistent with basic intuition, $S_N$ is a decreasing function of $N$ and
also decreases as $z\to 1$ with $N$ fixed.  It should be emphasized that
Eq.~(\ref{PN-asymp}) applies in the limit of $z\sqrt{\ln N}\gg 1$, which is
extremely hard to achieve by direct simulation.  For example, if the lamb
starts halfway between the haven and the lions ($z=\frac{1}{2}$), then for
$N=10^4$, the argument of the complementary error function is $z\sqrt{\ln
  N}\approx 1.52$; for $N=10^{16}$, $z\sqrt{\ln N}\approx 3.03$.  Conversely
to reach $z\sqrt{\ln N}=10$ requires $N=e^{400}\approx 10^{174}$.  Notice
also that Eq.~(\ref{PN-int}) matches the survival probability given by
Eq.~(\ref{fza}) for a ballistically-moving boundary when $N$ reaches a
critical value for which the completion time $t^*=L^2/A$ also equals $L/v$.

More rigorously, we should also incorporate the absorbing boundary condition
at $L(t)$, corresponding to the lamb getting eaten by the closest lion.  This
problem of a fixed absorbing boundary at $x=0$ and a moving absorbing
boundary at $x=L-\sqrt{At}$ does not seem readily soluble, however.  Instead,
we investigate a related model in which the boundary motion mimics that of
the closest lion, but is engineered to be soluble.  As we shall show, the
ultimate survival probability for this alternative problem has a
qualitatively similar dependence on system parameters as Eq.~(\ref{PN-asymp}).
Consider the toy model in which the closest lion coordinate is $L_{\rm
  toy}(t) = \sqrt{L^2\!-\! B t}$ (compared to $L(t)=L\!-\!\sqrt{At}$, with
$A=4D\ln N$, for $N\gg 1$ diffusing lions).  These two boundaries satisfy the
inequality $L_{\rm toy}(t)>L(t)$ and both reach the origin at the same time
when $B=A$.  Thus the toy model remains an upper bound for the true survival
probability.

It is again convenient to treat the evolution of the system in the
two-dimensional space $(x,L)$.  Let $S(x,L)$ be the probability that the lamb
successfully reaches the haven, where $x$ and $L$ denote the initial
positions of the lamb and the boundary respectively.  Following the same
approach as in Sec.~\ref{inf}, we write the backward equation for $S(x,L)$.
In a small time interval $[0,\Delta t]$ the lamb moves to $x+\eta(0)\Delta
t$, where $\eta(t)$ is Gaussian white noise with zero mean, and the boundary
moves to $L - (B/2L) \Delta t$, where $B/2L$ is the boundary speed.  The
survival probability now satisfies $S(x,L)= \langle S(x+\eta(0) \Delta t,
L-(B/2L)\Delta t)\rangle$, and expanding the right-hand side to lowest order
gives the backward equation
\begin{equation}
\label{Stoy}
D \frac{d^2S}{dx^2} - \frac{B}{2L} \frac{dS}{dL} =0 
\end{equation}
for $0\le x\le L$, with the boundary conditions $S(0,L)=1$ and $S(L,L)=0$.
To solve (\ref{Stoy}) we make the scaling ansatz $S(x,L) = f(y)$, with
$y=\gamma z$, where $\gamma = \sqrt{B/(2D)}$ and $z=\frac{x}{L}$, and find
that the scaling function satisfies
\begin{equation}
\label{f}
f'' + y\, f' =0\,,
\end{equation}
for $0\le y\le 1$, with the boundary conditions $f(0)=1$ and $f(\gamma)=0$;
here the prime denotes differentiation with respect to $y$.  Integrating once
gives $f\propto e^{-y^2/2}$, and integrating again gives
\begin{equation}
\label{f-sol}
  f(z) = 1- \mathrm{erf}(\gamma z/\sqrt{2})/\mathrm{erf}(\gamma/\sqrt{2})\,, 
\end{equation}
where the constants are determined by the boundary conditions.  Substituting
in $\gamma = \sqrt{B/(2D)}$ and $B=4D\ln N$ gives the asymptotic behavior
\begin{equation}
  f(z) \approx [N^{-z^2} - N^{-1}]\sim N^{-z^2}~.
\end{equation}
This upper bound has the same asymptotic behavior as (\ref{PN-asymp}) and
suggests that the heuristic approach should be quite accurate.

\section{Simulations}
\label{sim}

We now present simulation results for the lamb-lion-haven system.  While a
direct simulation is simple to code, it becomes prohibitively slow when $N$
is large.  We have therefore developed two complimentary approaches to
determine the survival probability in the large-$N$ limit.

\subsection{Probability Propagation}
\label{propagation}

Probability propagation is well-suited for probing the case of $N\gg 1$,
where we replace the position of the closest lion by a deterministic
absorbing boundary, $L(t)$, that moves according to Eq.~(\ref{lstar}).  Here,
the constant $A$ can be chosen as the mean or most probable position of the
closest lion or any other reasonable positional metric.  We choose to set
$A=4D\ln{N}$, which is the leading behavior in Eq.~(\ref{lstar}).  The
omission of higher-order corrections, which slightly decrease $A$, lead to a
more slowly-moving boundary and a correspondingly slightly larger survival
probability.  Thus probability propagation should provide a lower bound to
the true survival probability.

Let $P(x,t)$ be the probability that the lamb is at $x$ at time $t$.  At each
time step, the probability in the interior region $2<x<\lfloor L(t)\rfloor
-1$ propagates according to
$P(x,t+1)=\frac{1}{2}P(x-1,t)+\frac{1}{2}P(x+1,t)$; here $\lfloor
L(t)\rfloor$ is the largest integer less than $L(t)$.  At the edge sites
$P(1,t+1)=\frac{1}{2}P(2,t)$ and $P(\lfloor
L(t)\rfloor,t+1)=\frac{1}{2}P(\lfloor L(t)\rfloor -1,t)$.  Probability
elements that reach either $x=0$ or $\lfloor L(t)\rfloor+1$ do not propagate
further and remain in place.  Probability propagation continues until $L(t)$
reaches $x=0$.  The total probability at $x = 0$ at this termination time
gives the survival probability of the lamb.

We used probability propagation to obtain $S_N(z)$ for $N$ up to $10^{500}$.
We used quadruple precision variables to ensure accuracy of the probability
values throughout the propagation.  The initial value of $L$ was chosen to be
the smallest such that finite-size effects were imperceptible --- this ranged
from $L =1000$ for small $N$ to $L = 30,000$ for the largest $N$ values.

\subsection{Event-Driven Simulation}

A naive simulation is simply to move every lion and the lamb by $\pm 1$ at
each time step, an approach which is prohibitively slow for large $N$.
However, there is no need to simulate every single random-walk step,
particularly if the lamb is far from both the haven and nearest lion.  This
motivates using an event-driven simulation, in which we propagate all
particles over a time that corresponds to a finite fraction of the time
needed for a reaction to actually occur --- either the lamb reaching the
haven or getting eaten by the closest lion.

Let $y$ be the minimum of the distances between the lamb and the nearest
lion, and between the lamb and the haven.  We could move every particle
according to a binomial distribution of ${y}/{2}-1$ steps because there is no
possibility that the lamb meets any of the lions or reaches the haven during
this update.  However, this approach is unnecessarily stringent because each
particle moves a typical distance that is only of the order of $\sqrt{y}$.
Thus we increment the number of steps by $m$, where
\begin{equation}
\label{update}
  m= \cases{  y^2/2Y & if $y\geq Y$, \cr
   {y}/{2} & if $y<Y$, }
\end{equation}
and move every particle according to a binomial distribution of $m$ steps.
Note that these update rules match at the crossover separation $y=Y$.  After
each such update, we check if the lamb has reached or crossed over the
position of the haven or that of any lion, in which case the simulation is
finished.

For $y<Y$, $m=\frac{y}{2}$ and the lamb cannot reach either the haven or any
lion during the update; this part of the simulation is exact.  For $y\geq Y$,
there is a non-zero probability that the lamb trajectory could cross the
haven or a lion trajectory and then cross back during the update.  However,
by choosing $Y$ appropriately, the probability of error due to such crossing
trajectories can be made vanishingly small.  We found that $Y=15$ gave an
excellent compromise between accuracy and efficiency.  We also checked that
simulations results with the update rule (\ref{update}) are essentially
identical with exact results that arise by choosing $Y=\infty$ in the update
rule (\ref{update}).

\subsection{Results}

\begin{figure}[ht]
\center{
\includegraphics[width=0.5\textwidth]{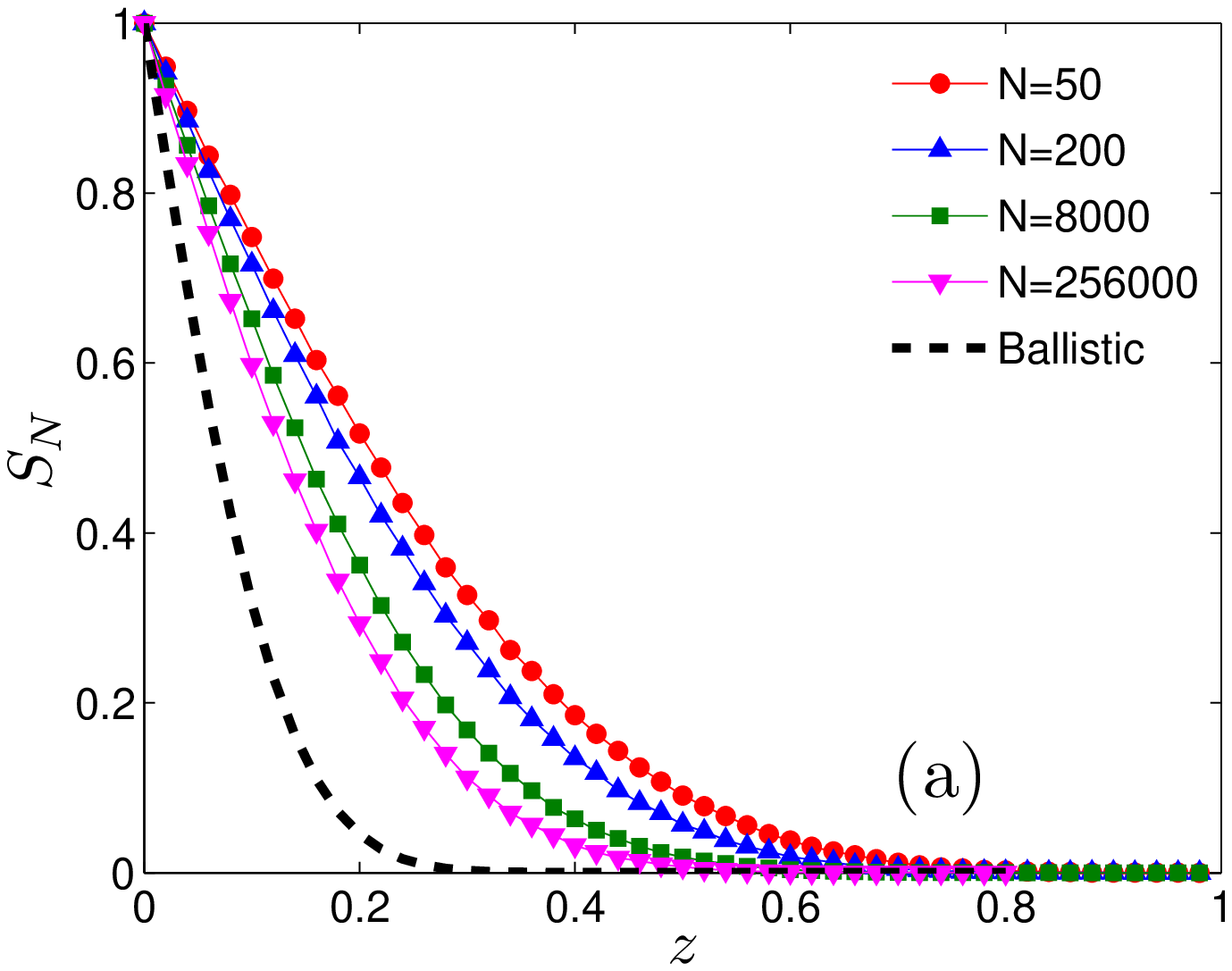}\includegraphics[width=0.5\textwidth]{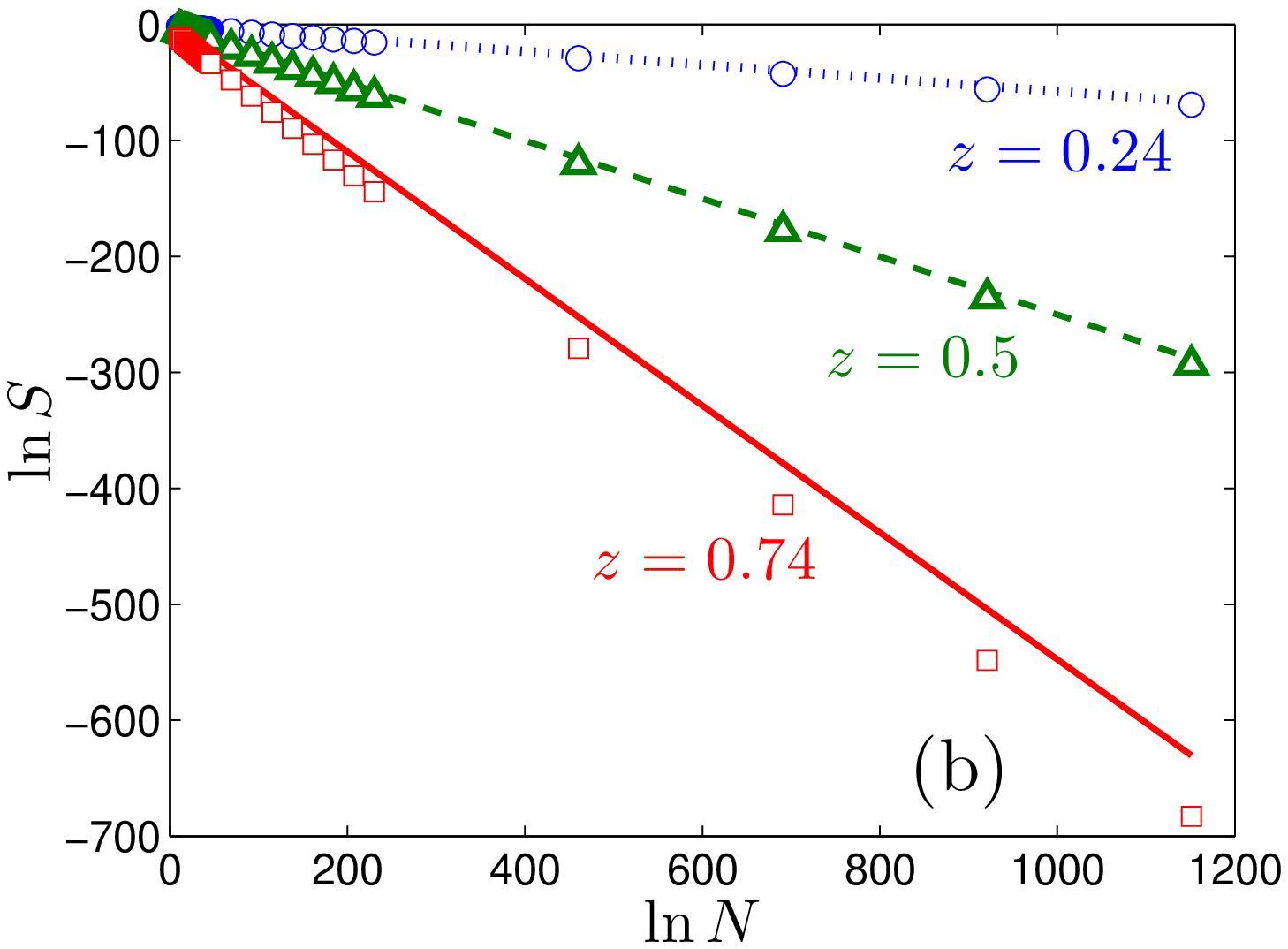}
\caption{(a) Dependence of the survival probability on $z$ for representative
  values of $N$ from event driven simulations. Dashed curve is analytic
  solution for ballistic lion motion from Eq.~(\ref{fz}) where $v=1$,
  $D=1/2$, and $L=100$. (b) Survival probability versus number of lions $N$
  for three representative $z$ values.  Curves give the analytic prediction
  $S=N^{-z^2}$ and the symbols represent probability propagation results.}
\label{Svsz}
}
\end{figure}

In Fig.~\ref{Svsz}(a) we show the dependence of the ultimate survival
probability versus scaled initial position $z=\frac{x}{L}$ for $N$ up to
$256,000$, with $10^5$ realizations for each data point, from event-driven
simulations.  For $N\gg 1$, these survival probabilities gradually converge,
as $N$ increases, to a limiting curve that corresponds to the system where
the last lion moves ballistically.  The lions are all initially at $L=100$
and we verified that the survival probability depends only on the ratio
$x/L$, without any explicit finite-$L$ dependence.  This independence on $L$
emerges when $L\geq 100$ and thus we focus on the smallest system ($L=100$)
where finite-size effects are negligible.

We also examined the dependence of $S_N$ on $N$ for fixed $z$ to test the
asymptotic power-law behavior $S\sim N^{-z^2}$ of Eq.~(\ref{PN-asymp}).  Our
analytical prediction matches the simulation quite well for $z\lesssim 0.5$
(Fig.~\ref{Svsz})(b).  However, a small but slowly growing discrepancy
arises as $z$ is increased beyond 0.5.  The source of this discrepancy is
that the heuristic derivation of Sec.~\ref{general} ignores the existence the
absorbing boundary caused by the last lion.  When $z$ approaches 1, the lamb
starts sufficiently close to the last lion that the assumption of ignoring
the boundary caused by the last lion is no longer valid.

\begin{figure}[ht]
\center{
\includegraphics[width=0.5\textwidth]{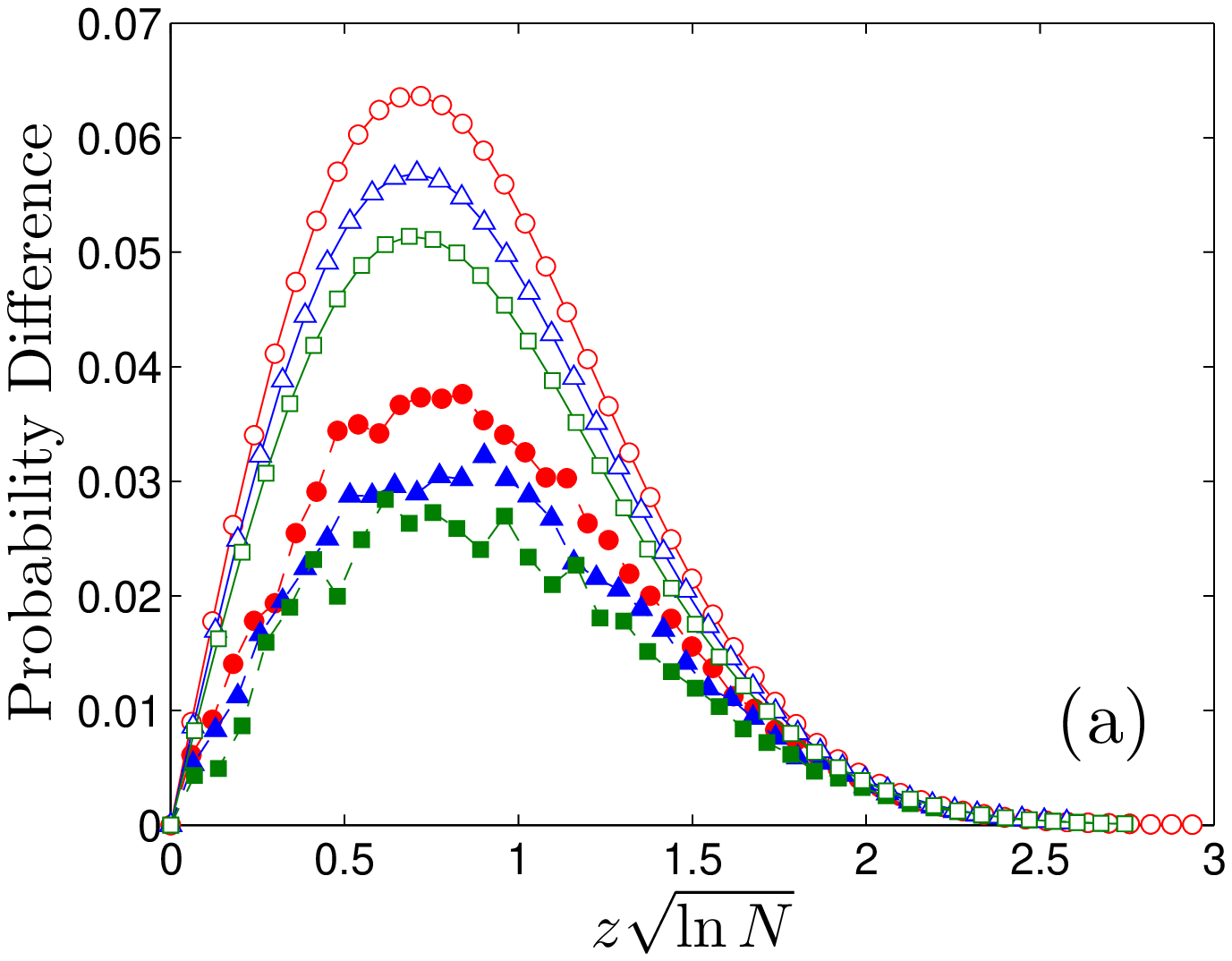}\includegraphics[width=0.5\textwidth]{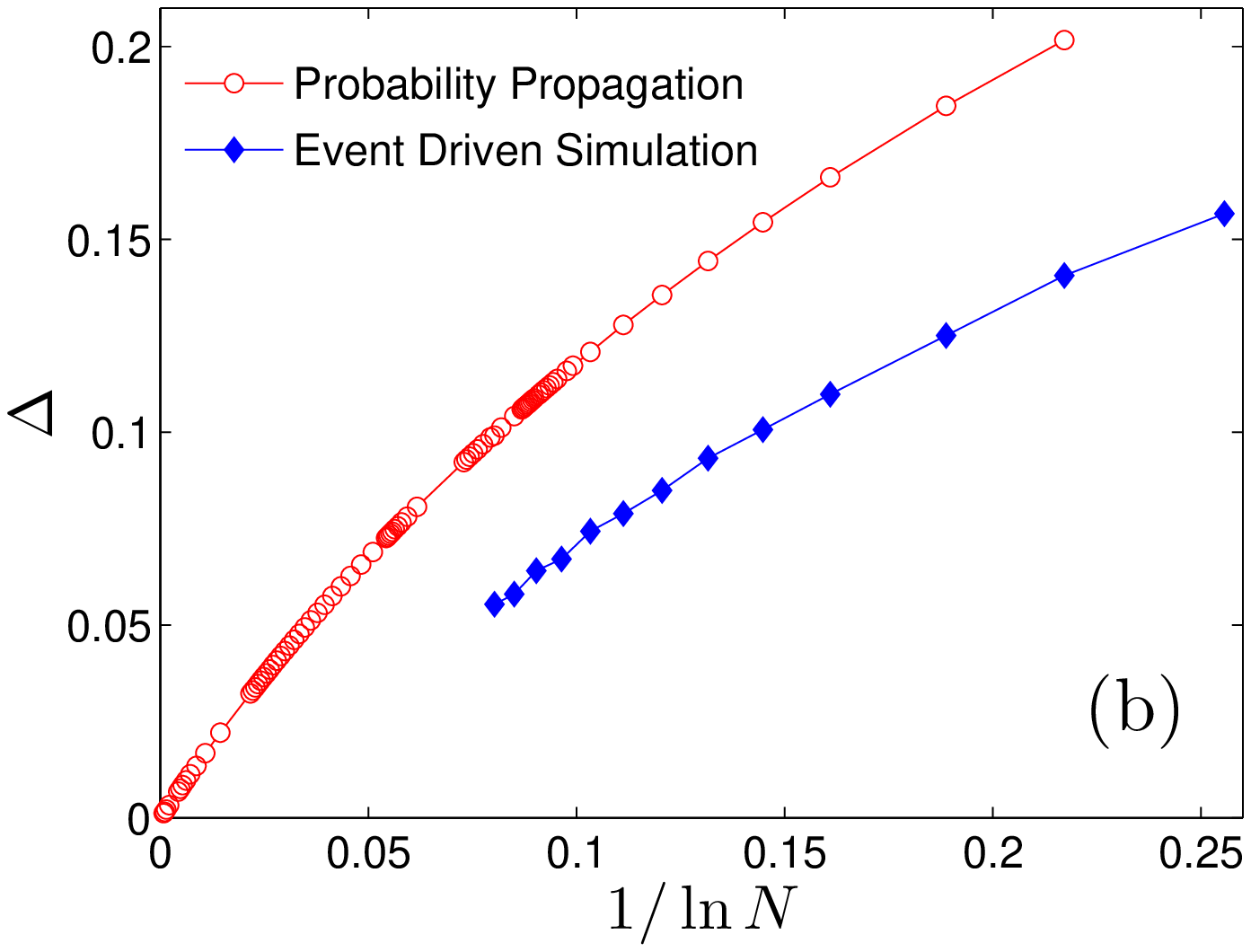}
\caption{(a) Difference between simulated survival probability and
  $S_N(z)=\mathrm{erfc}(z\sqrt{\ln{N}})$ as a function of $z$ for various
  $N$.  Open symbols correspond to probability propagation while filled
  symbols correspond to event-driven simulations.  Circles correspond to
  $N=8,000$, triangles to $N=32,000$, and squares to $N=128,000$. (b)
  Relative area difference $\Delta$ versus $1/\ln{N}$ for probability
  propagation ($\circ$) and event-driven ($\blacklozenge$) simulations.}
\label{simDiff}
}
\end{figure}

Finally, we compare our two simulation approaches with each other and with
our our heuristic prediction $S_N\lesssim\mathrm{erfc}(z\sqrt{\ln{N}})$ from
Eq.~(\ref{PN-int}).  By construction, the event-driven simulation is more
accurate because it explicitly follows the stochastic motion of the lamb and
the lions.  Our heuristic prediction (\ref{PN-int}) provides an upper bound
for large $N$, but this regime is not feasible to simulate with the
event-driven algorithm.  Conversely, the probability propagation simulation
can be implemented for arbitrarily large $N$ but suffers from systematic
error because it assumes the closest lion position to be deterministic. 

Figure~\ref{simDiff}(a) illustrates the convergence of the simulation
results to Eq.~(\ref{PN-int}) where the difference between the simulated
value of $S_N(z)$ and $\mathrm{erfc}(z\sqrt{\ln{N}})$ is plotted as a
function of $z$ for representative $N$ values.  We quantify this difference
by $\Delta\equiv (A_{\rm a}\!-\!A_{\rm s})/A_{\rm a}$, where $A_{\rm a}=\int
S_N(z)dz$, with $S_N(z)=\mathrm{erfc}(z\sqrt{\ln{N}})$, is the area beneath
the analytic survival curve and similarly for the area beneath the simulated
curve.  Figure~\ref{simDiff}(b) shows that $\Delta\to 0$ as $N\to\infty$
for the probability propagation algorithm.  A similar, but not identical
convergences arises in the event-driven simulation, but the method cannot
reach the large-$N$ regime.  These results provide strong evidence that the
survival probability is indeed given by
$S\rightarrow\mathrm{erfc}(z\sqrt{\ln{N}})$ as $N\rightarrow\infty$.

\section{Outlook}
\label{conc}

The presence of a haven adds an intriguing element to the classic capture
process of a single lamb in the presence of $N$ diffusing lions.  Now the
basic question is whether the lamb can reach safety at the haven before it is
eaten by one of the lions.  We investigated the dependence of the ultimate
survival probability of the lamb, $S_N(x,L)$, on the number of lions $N$ and
also on the initial positions of the lamb ($x$) and the lions (all at $L$,
for simplicity).  By a rough heuristic argument, we found that $S_N$ has the
asymptotic behavior $S_N\lesssim \mathrm{erfc}(z\sqrt{\ln N})$, and this
function has the unusual leading behavior $S_N \sim N^{-z^2}$, where $z=x/L$.
It is remarkable that a simplistic approach gives such an unusual and rich
result.  However, the approach to this asymptotic regime is extremely slow
and it is necessary to simulate a system that corresponds to $N$ of the order
of $10^{500}$ lions before the asymptotic behavior becomes apparent.

It is natural to ask about the properties of the ultimate survival
probability in higher dimensions.  For diffusive capture in an unbounded
system, the case of one dimension is the most interesting.  However, the
presence of a haven now makes the higher-dimensional problem nontrivial.
For example, in two dimensions, a natural setting would be a diffusing prey,
$N$ diffusing predators, and a circular haven of radius $R$ centered at the
origin.  Because of the recurrence of diffusion in two dimensions, the prey
will eventually reach the haven if there are no predators, but the mean time
to reach the haven is infinite.  What happens when predators exist?  How does
the survival probability depend on the number of predators and on the initial
positions of the prey and predators?  How long does it take for the capture
process to end?  Another interesting two-dimensional geometry is a
semi-infinite planar haven.  Finally, in three dimensions, the transience of
diffusion could lead to very different properties for the survival
probability than in two dimensions.

\medskip We thank Paul Krapivsky for helpful discussions and advice.  AG,
NKP, and SR thank NSF grant DMR-0906504 for partial financial support of this
research.  S.N.M. acknowledges partial support by ANR grant 2011-BS04-013-01
WALKMAT and by the Indo-French Centre for the Promotion of Advanced Research
under Project 4604-3.


\end{document}